\documentclass[aps,prb,showpacs,preprint,groupedaddress,floatfix]{revtex4-1}

\usepackage{graphicx}
\usepackage{epstopdf}
\usepackage{multirow}
\usepackage[%
  pdfstartview=={FitH},
  pdftitle={Film growth of Mn3O4(001) on Ag(001)},
  pdfauthor={Konrad Gillmeister},%
  bookmarks=true,
  hyperfigures=true,
  unicode=true,
  colorlinks=true,
  linkcolor=black,	
  citecolor=black,	
  urlcolor=black,	
]{hyperref}  

\begin{document}

\title{Mn$_3$O$_4$(001) film growth on Ag(001) - a systematic study using NEXAFS, STM, and LEED}
\author{K. Gillmeister}
\affiliation{Institut f\"{u}r Physik, Martin-Luther-Universit\"{a}t Halle-Wittenberg, Halle, Germany}
\author{M. Huth}
\affiliation{Institut f\"{u}r Physik, Martin-Luther-Universit\"{a}t Halle-Wittenberg, Halle, Germany}
\altaffiliation[Now at: ]{Max-Planck-Institut f\"ur Mikrostrukturphysik, Weinberg 2, D-06120 Halle, Germany}
\author{R. Shantyr}
\affiliation{Institut f\"{u}r Physik, Martin-Luther-Universit\"{a}t Halle-Wittenberg, Halle, Germany}
\author{M. Trautmann}
\affiliation{Institut f\"{u}r Physik, Martin-Luther-Universit\"{a}t Halle-Wittenberg, Halle, Germany}
\author{K. Meinel}
\affiliation{Institut f\"{u}r Physik, Martin-Luther-Universit\"{a}t Halle-Wittenberg, Halle, Germany}
\author{A. Chass\'{e}}
\affiliation{Institut f\"{u}r Physik, Martin-Luther-Universit\"{a}t Halle-Wittenberg, Halle, Germany}
\author{K.-M. Schindler}
\affiliation{Institut f\"{u}r Physik, Martin-Luther-Universit\"{a}t Halle-Wittenberg, Halle, Germany}
\email{karl-michael.schindler@physik.uni-halle.de}
\author{H. Neddermeyer}
\affiliation{Institut f\"{u}r Physik, Martin-Luther-Universit\"{a}t Halle-Wittenberg, Halle, Germany}
\author{W. F. Widdra}
\affiliation{Institut f\"{u}r Physik, Martin-Luther-Universit\"{a}t Halle-Wittenberg, Halle, Germany}
\affiliation{Max-Planck-Institut f{\"u}r Mikrostrukturphysik, Halle, Germany}
\date{\today}

\begin{abstract}
The film growth of Mn$_3$O$_4$(001) films on Ag(001) up to film
thicknesses of almost seven unit cells of Mn$_3$O$_4$ has been
monitored using a complementary combination of near-edge X-ray
absorption fine structure spectroscopy (NEXAFS), scanning tunneling
microscopy (STM), and low-energy electron diffraction (LEED).  The
oxide films have been prepared by molecular beam epitaxy.  Using
NEXAFS, the identity of the Mn oxide has clearly been determined as
Mn$_3$O$_4$.  For the initial stages of growth, oxide islands with
p(2$\times$1) and p(2$\times$2) structures are formed, which are
embedded into the substrate.  For Mn$_3$O$_4$ coverages up to 1.5 unit
cells a p(2$\times$1) structure of the films is visible in STM and
LEED. Further increase of the thickness leads to a phase transition of
the oxide films resulting in an additional c(2$\times$2) structure
with a 45$^\circ$ rotated atomic pattern.  The emerging film
structures are discussed on the basis of a sublayer model of the
Mn$_3$O$_4$ spinel unit cell.  While the polarity of the island edges
determines the structure of initial islands, the surface energy of
thicker layers is remarkably reduced by a film restructuring.
\end{abstract}

\pacs{}

\maketitle

\section{Introduction}

Building up on the seminal investigations of oxide films of NiO
\cite{Hannemann1994} and CoO \cite{Sebastian1998}, manganese oxide
(MnO) films have been grown and investigated on Pt(111)
\cite{Rizzi2001, Hagendorf2008, Sachert2010, Martynova2013}, Pd(001)
\cite{Allegretti2007, Franchini2009PRB, Franchini2009JCP, Li2009},
Ru(001) \cite{Nishimura2000}, Rh(111) \cite{Zhang2012}, and Ag(001) \cite{Mueller2002,
Nagel2007SS, Nagel2007PRB, Chasse2008, Soares2005}.  A common feature
of all these film systems is that the rock salt unit cells of these
materials have only one sort of atomic sublayer in the (001)-plane.
Here, we focus on Hausmannit (Mn$_3$O$_4$) a spinel
with a stack of 8 sublayers and a Jahn-Teller distortion of the unit 
cell in c direction.  All its simple bulk terminations in the (001)-plane 
are polar, which makes it particularly
interesting to study the layer-by-layer build up of Mn$_3$O$_4$ films.

\begin{figure}
  \includegraphics[height=7.5cm]{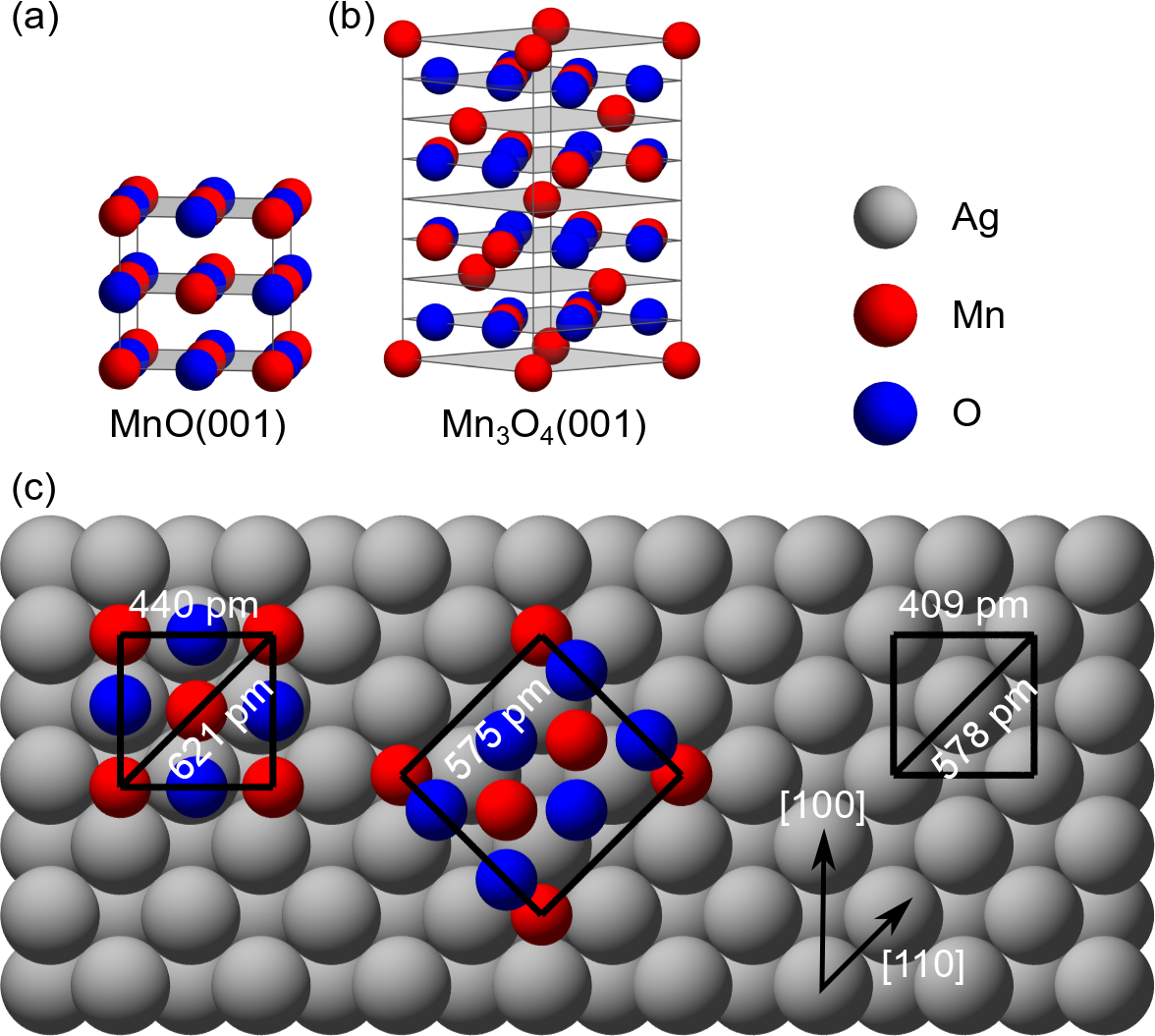}
	\caption{Hard sphere models of unit cells of (a) rock salt MnO(001)
	and (b) spinel Mn$_{3}$O$_{4}$(001) on Ag(001).  (c) Lattice
	fittings for MnO(001)/Ag(001) (lattice misfit 7.6\%) and 45$^\circ$
	rotated Mn$_{3}$O$_{4}$/Ag(001) (lattice mismatch
	0.3\%).}\label{fig:UnitCell}
\end{figure}

So far, only the growth of Mn$_3$O$_4$(110) on SrTiO$_3$(110)
\cite{Gorbenko2002} and the growth of Mn$_3$O$_4$(001) on
MnO(001) \cite{Bayer2007} are reported.  For
Mn$_3$O$_4$(001)/MnO(001), electron diffraction showed that the
Mn$_3$O$_4$(001) films grow parallel to $\langle 110\rangle$
directions of the underlying MnO(001) lattice.  This means that the
($a_{Mn_3O_4} \times a_{Mn_3O_4}$) unit mesh is rotated by 45$^\circ$
resulting in a ($2 \times 2$) LEED pattern with respect to the
MnO(001) substrate.  Such a 45$^\circ$ rotation has also been observed
for other film - substrate combinations ,e.g., LiF, KCl, and NaI on
MgO(001) \cite{Takayanagi1978} and NaCl on Ag(001) \cite{Pivetta2005}.
According to the 'roles of lattice fitting in epitaxy'
\cite{Takayanagi1978}, a 45$^\circ$ rotated orientation has to be
expected if the ratio $a_{film}\,:\,a_{substrate}$ of the lattice
parameters is close to $\sqrt{2}$.  For the 45$^\circ$ rotated growth
of Mn$_3$O$_4$(001) on Ag(001) ($a_{Ag}$ = 409~pm, $a_{Mn_3O_4}$ =
576~pm) one obtains an almost vanishing misfit of 0.3\% (Fig.
\ref{fig:UnitCell}).  Consequently, one can expect a perfect film
structure when growing Mn$_3$O$_4$(001) films on Ag(001) compared to
MnO(001) films on this substrate.

In the present study, Mn$_3$O$_4$ films have been prepared on Ag(001)
by means of reactive Mn deposition in O$_2$ atmosphere
\cite{Marre1993}.  Structure, morphology, composition, and electronic
properties of the films have been characterized using low-energy
electron diffraction (LEED), scanning tunneling microscopy (STM),
Auger electron spectroscopy (AES), and near edge X-ray absorption fine
structure spectroscopy (NEXAFS).  Starting from the initial island
stage up to thicknesses of several Mn$_3$O$_4$ unit cells (height:
946~pm), the development of the surface structure as a function of
film thickness and annealing is reported.

\section{Experimental}
Experiments have been performed in three different ultra-high vacuum
(UHV) systems.  The first is equipped with an STM, a spot profile
analysis (SPA)-LEED optics, and a cylindrical mirror analyzer (CMA)
for Auger electron spectroscopy (AES).  The second system contains a
conventional LEED optic and a low-temperature STM, operating at 100~K.
The base pressures of both chambers are in the low 10$^{-10}$~mbar
range.  In addition, a third UHV chamber equipped with a LEED optics
has been used for NEXAFS studies recorded at the beamline UE56-2 PGM-2
(energy range: 100-1000~eV) at the synchrotron radiation facility
BESSY II \cite{Sawhney1997}.  Spectra were recorded in total electron
yield mode.  Normalization to the incident X-ray flux was achieved
according to the photo current from the last refocusing mirror of the
beamline.  All presented LEED patterns have been recorded with the
SPA-LEED optics.

The Ag(001) crystals (miscut $< 0.2^\circ$) were cleaned by cycles of
Ar\textsuperscript{+} ion sputtering (600~V, 2~$\mu$A) at room temperature and
subsequent heating at 630~K until they showed a clean, defect free
surface in STM and sharp spots in the LEED pattern.  Manganese was
evaporated from Ta crucibles heated by electron bombardment.  The
deposition rate of 30~pm (i.e. 0.03~ML) per minute was calibrated by
means of a quartz microbalance, AES, LEED, and STM and controlled by
monitoring the flux of Mn ions as reported previously
\cite{Sachert2010}.  The deposition was performed at
sample temperatures between room temperature and 450~K.

The integral structure of the as grown Mn$_3$O$_4$ films was
characterized by LEED. The local film morphology was investigated by
STM in constant current mode.  The equivalence of films obtained in
the different UHV systems was checked using LEED.

\section{Results}

\subsection{NEXAFS of MnO and Mn$_3$O$_4$}
Manganese can occur in various oxidation states.  In our case, the
particular Mn oxidation state is determined by the oxygen partial
pressure and substrate temperature during growth.  At low oxygen
partial pressures MnO with Mn$^{2+}$ should be formed.  With
increasing pressure, a transition to the formation of the next higher
oxide Mn$_3$O$_4$ with a mix of Mn$^{2+}$ and Mn$^{3+}$ is expected.
In the following, LEED and NEXAFS are used to identify the films
prepared at different oxygen partial pressures.  MnO(001) films on
Ag(001) with a thickness of 4~ML have been obtained by reactive
evaporation of Mn in $5 \times 10^{-8}$~mbar O$_2$ with the Ag(001)
substrate at room temperature.  Annealing the film results in
brilliant and sharp LEED spots showing the ($1 \times 1$) pattern of
strained MnO(001) as reported previously \cite{Soares2005,
Chasse2011}.  Assuming pseudomorphic growth of the MnO film on the
Ag(100) substrate, 4~ML MnO equal to 48.0 Mn ions/nm$^2$.

\begin{figure}
  \includegraphics[scale=0.5]{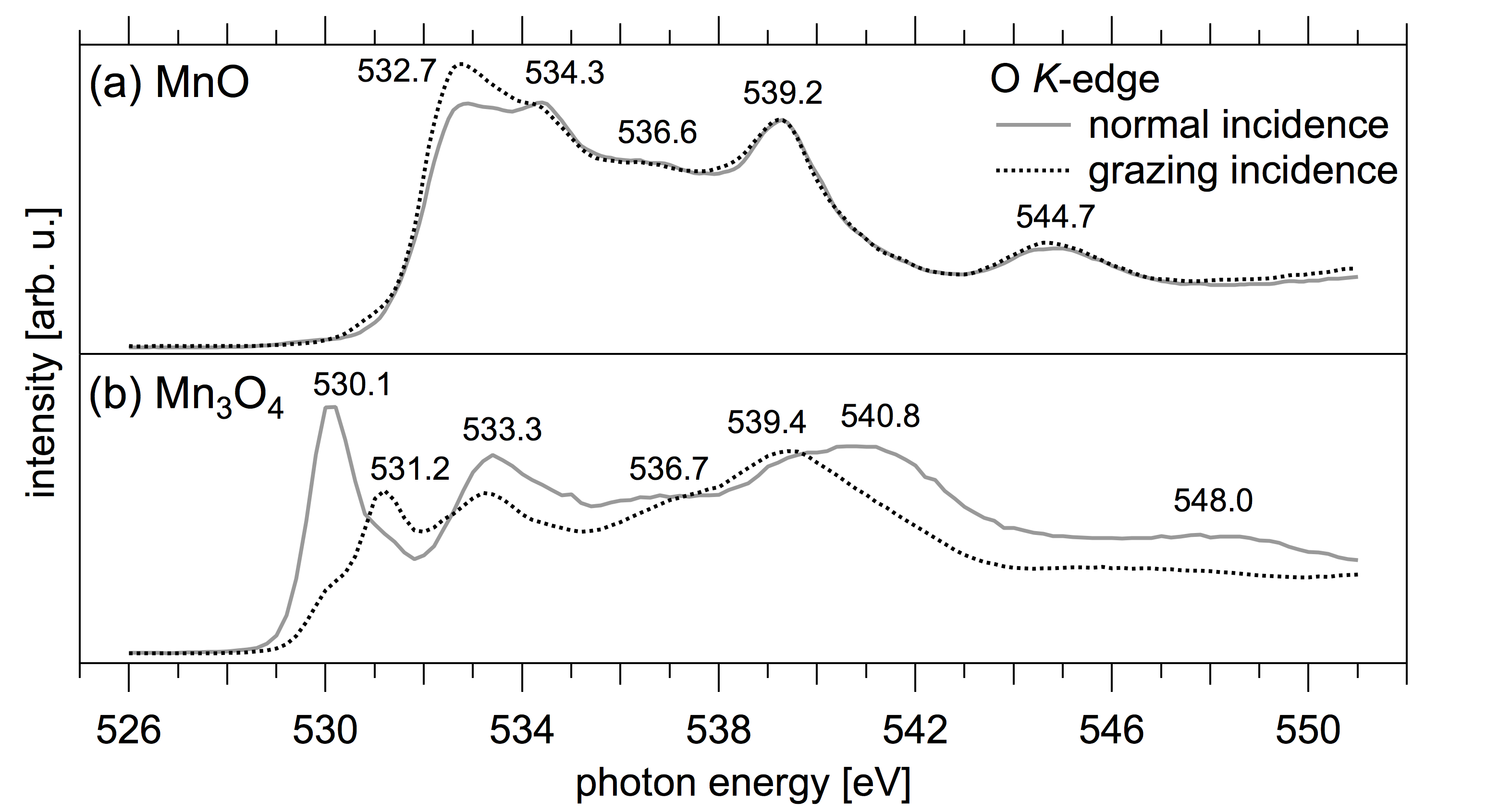}
  \caption{NEXAFS spectra of (a) 4~ML MnO and (b) 20~MLE
    Mn$_3$O$_4$ on Ag(001) for normal and grazing X-ray
    incidence.  For a detailed discussion see the text.
  }\label{fig:NEXAFS}
\end{figure}

NEXAFS spectra at the O K absorption edge are
depicted in Fig.  \ref{fig:NEXAFS}a.  The spectra 
agree with published spectra of MnO films grown on
Ag(001) \cite{Nagel2007PRB} and are characteristic for MnO(001).  
They show only minor differences 
between normal (0$^\circ$) and grazing light incidence (70$^\circ$
off-normal).  This indicates only minor deviations from a cubic
bulk-like structure of MnO (Fig.  \ref{fig:UnitCell}).  In the bulk
structure, the octahedral coordination of the Mn ions by six O ions
leads to an isotropic environment and no NEXAFS dependencies on angle
of light incidence or polarization.  In ultrathin films, however, MnO
growths strained due to the lattice mismatch between Ag and MnO. For
a thickness of 4~ML the film is not yet relaxed to the bulk structure,
but exhibits a lateral compression and a corresponding vertical
expansion \cite{Chasse2011}.  The resulting lowering of the symmetry
from cubic to tetragonal might explain the small differences in
Fig.  \ref{fig:NEXAFS} at 533~eV. In addition, the reduced
coordination of Mn ions in the top surface layer and in the bottom
layer at the interface to the Ag substrate can contribute to this
anisotropy.

In order to obtain the oxygen richer Mn$_3$O$_4$ films, the O$_2$
pressure during Mn evaporation was increased systematically. 
At an O$_2$ pressure of $5 \times 10^{-7}$~mbar, a change in the LEED
pattern and the NEXAFS spectra is observed.  The LEED spots 
become sharp and brilliant again.  Depending on film thickness, 
a p($2 \times 1$) or an apparent ($2 \times 2$) superstructure relative to the
Ag(001) substrate develops.  The latter is composed of p($2 \times
1$) and c($2 \times 2$) domains, as will be shown later.  The
superstructures observed correspond to a pseudomorphic Mn$_3$O$_4$
film rotated by 45$^\circ$ around the surface normal as visualized schematically
in Fig.  \ref{fig:UnitCell}.

The corresponding NEXAFS spectra for a Mn$_3$O$_4$(001) film with a 
thickness equivalent to 20 monolayers (MLE, 1 MLE $\widehat{=}$ 12 
Mn ions/nm$^2$) 
at the O K-edge in Fig.  \ref{fig:NEXAFS}b are very
similar to those of bulk Mn$_3$O$_4$ \cite{Gilbert}.  The film
thickness is given in terms of the equivalent amount of Mn ions, which
makes the comparison of different phases easier. 
The bulk crystal structure of Mn$_3$O$_4$ has a unit cell height of 947,0 pm 
consists of 8 sublayers. A unit cell high Mn$_3$O$_4$ film corresponds to 36 Mn
ions/nm$^2$.  The Mn$_3$O$_4$ film of fig.  \ref{fig:NEXAFS}, therefore, has a thickness of 6.7
unit cells or about 50 sublayers. The NEXAFS spectrum of the
Mn$_3$O$_4$ film shows 7 resonances at 530.1~eV, 531.2~eV, 533.3~eV, 
536.7~eV, 539.4~eV, 540.8, and 548.0 eV.
The energies of all resonances agree well 
with resonances in electron loss fine structure spectra at the O K-edge of Mn$_3$O$_4$ \cite{Laffont2010,
Chen2014}.  The presence of $\alpha$-Mn$_2$O$_3$ can clearly be ruled
out, since its characteristic resonance at 542~eV is missing in \ref{fig:NEXAFS}b. Instead, the two resonances at 539.4 and
540.8~eV and the one at 548~eV are characteristic for
Mn$_3$O$_4$ and not present at all in $\alpha$-Mn$_2$O$_3$.  However, the
spectra of $\gamma$-Mn$_2$O$_3$ and Mn$_3$O$_4$ are very similar and
can hardly be distinguished \cite{Kim2005}.  This similarity
originates in nearly identical geometric structures with no difference
in bond lengths and angles.  The only difference is the occupation of
octahedral and tetrahedral sites by Mn ions and $\gamma$-Mn$_2$O$_3$
can be considered as a vacancy structure of Mn$_3$O$_4$, whereby 
the occupation of octahedral or tetrahedral sites has not fully
been resolved.  Investigations in the late 50's arrived at
controversial results (tetrahedral vacancies by Goodenough et al.
\cite{Goodenough1955} and octahedral vacancies by Sinha et al.
\cite{Sinha1957}) which have not been further investigated later on.
Our subsequent discussion will be based on the vacancy-less structure
of Mn$_3$O$_4$, because we started from growth conditions
for MnO and slowly increased the oxygen pressure. Mn$_3$O$_4$ is
expected to be formed prior to Mn$_2$O$_3$ due to its lower 
oxygen content.

Compared to the NEXAFS spectra of MnO, the absorption onset in the Mn$_3$O$_4$
spectra is shifted characteristically to a lower photon energy.
Since this edge is very prominent, its absence in the first spectrum
clearly rules out even small amounts of Mn$_3$O$_4$ in the MnO film.

Contrary to MnO, the NEXAFS spectra of the Mn$_3$O$_4$
film strongly depend on the angle of light incidence.  Taking into account
that the bulk structure of Mn$_3$O$_4$ has a tetragonal unit cell
(Fig.  \ref{fig:UnitCell}) with different atomic arrangements of O
ions around Mn ions, the dependence of the x-ray absorption on the
angle of incidence indicates directly a c-axis alignment of the Mn$_3$O$_4$(001) film.
Since the shorter a axis of Mn$_3$O$_4$ matches
better to the lattice of the substrate, 
preferential alignment of the c-axis normal to the surface is assumed.

\subsection{Initial stages of film formation}

\begin{figure}
  \includegraphics[height=4.1cm]{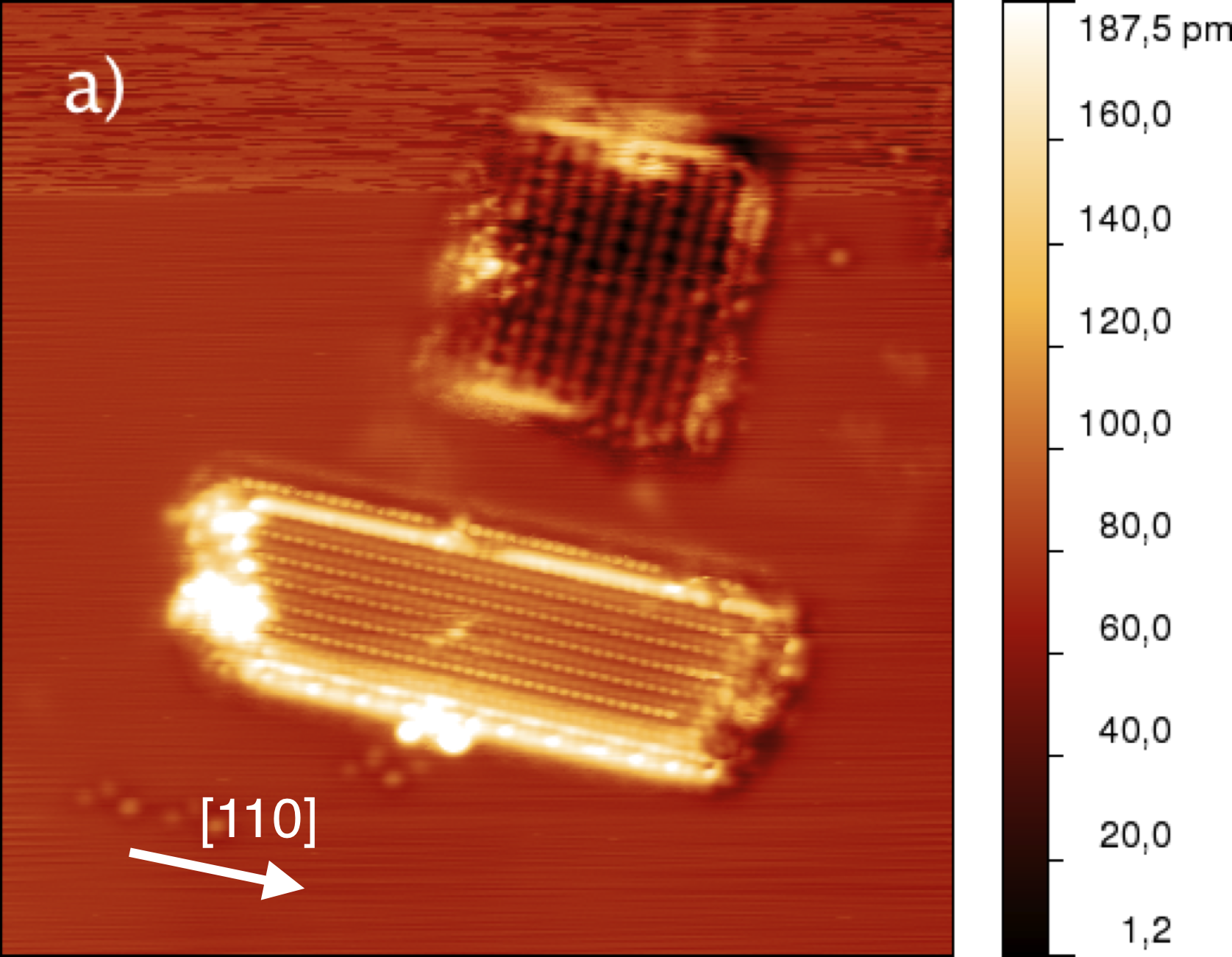}
  \includegraphics[height=4.1cm]{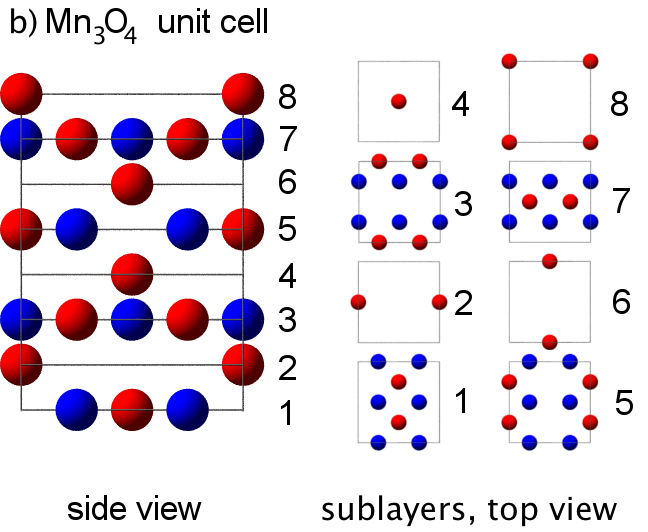}
  \includegraphics[height=4.1cm]{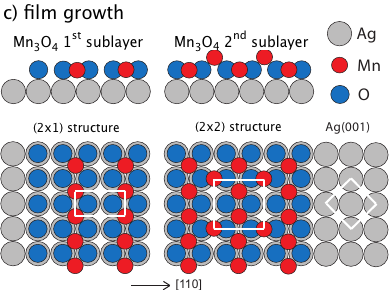}
  \caption{(a) STM image ($20 \times 20$~nm$^2$,
    -0.2~V, 0.7~nA) of a 0.1~nm thick Mn$_3$O$_4$ film on Ag(001) with a
    quadratic (top) and a rectangular (bottom) island, (b) sphere model
    of the Mn$_3$O$_4$ unit cell and its 8 sublayers, (c) film growth on
    Ag(001) for the first and second sublayers.
  }\label{fig:STM1} 
\end{figure}

The initial stages of film growth for Mn$_3$O$_4$ on Ag(001)
at a substrate temperature of 450 K have been characterized by STM. After the
deposition of 0.1~nm Mn$_3$O$_4$, two different island types can be
distinguished in STM images as shown in Fig.  \ref{fig:STM1}(a).  One type of
islands is of rectangular shape and has a well-resolved striped atomic
pattern pointing along [110] directions.  The other type of islands
appears darker, i.e. has a lower apparent height and a quadratic
atomic pattern with edges along [110] directions.  For the interpretation
of the STM images we start with hard sphere
models of the bulk Mn$_3$O$_4$(001) unit cell as depicted in Fig.  \ref{fig:STM1}b. 
Any
Jahn-Teller distortion is neglected here as well as later.  The bulk
structure of Mn$_3$O$_4$(001) is composed of two types of layers
(sublayers), both with a square (a $\times$ a) unit cell (Fig.
\ref{fig:STM1}b).  One layer is a mixed oxygen/manganese layer with
Mn$_2$O$_4$ composition (layers 1, 3, 5, 7).  The other layer is
manganese only (layers 2, 4, 6, 8).  These layers are stacked
alternatingly whereby consecutive layers of the same type are laterally displaced by
a/2 and rotated by 90$^\circ$.  Eight layers form the complete unit
cell with an overall height of c~=~947~pm.

The layered stacking scheme of the Mn$_3$O$_4$(001) unit cell readily
hints to possible interface structures of Mn$_3$O$_4$(001) on Ag(001) in (see Fig.
\ref{fig:STM1}c).  The square (a $\times$ a) unit cell is assumed to
be arranged along [110] directions of the Ag(001) substrate.  As
mentioned above this orientation leads to a nearly perfect lattice
match between the Mn$_3$O$_4$(001) layer and the Ag(001) substrate.
In Fig.  \ref{fig:STM1}c, a Mn$_2$O$_4$ sublayer (layer 1) is assumed
as the interface layer of the film.  Oxygen ions are located at top
positions of the Ag atoms, whereas manganese ions reside in fourfold
hollow sites, as found with CoO on Ag(100) \cite{Schindler2009}.  On
top of this mixed manganese/oxygen interface layer, the manganese ions
of the second sublayer (layer 2) are in bridge positions above two
oxygen ions.  According to this scheme, the Mn ions in the mixed
manganese/oxygen correspond to octahedrally coordinated Mn of the bulk
and form a p($2 \times 1$) structure.  The Mn ions of the manganese
only layer correspond to the tetrahedrally coordinated Mn in the bulk
and form a p($2 \times 2$) structure.  \footnote {Both superstructures
are given with respect to the Ag substrate.  } These two structures
fit well to the two different Mn$_3$O$_4$ islands at the initial stage of growth as found by STM in
Fig.  \ref{fig:STM1}a.  The quadratic atomic pattern of the top island
is aligned along [110] directions and corresponds to the p($2 \times
2$) structure whereas the bottom island displays the characteristic
rows of the p($2 \times 1$) structure running along $\langle 110
\rangle$ or $\langle 1\overline10\rangle$.  As the nominal film
thickness is one tenth of a unit cell, one can assume that the local
thickness of the bottom p($2 \times 1$) island is just one sublayer
and that of the top p($2 \times 2$) island is two sublayers.

Interestingly, the p($2 \times 1$) islands prefer a rectangular shape
with the long sides parallel to Mn ($2 \times 1$) rows.  On the
contrary, the p($2 \times 2$) islands grow in a nearly quadratic
shape.  Obviously, edge energies or diffusion barriers of the p($2
\times 1$) islands differ strongly along $\langle 110 \rangle$ and
$\langle 1\overline10\rangle$.  This can actually be related to the
atomic structures according to Fig.  \ref{fig:STM1}c.  Along the Mn
($2 \times 1$) rows the island edges are formed by a couple of
complete Mn and O rows, respectively.  In the perpendicular direction
every second Mn ion is missing in the Mn rows.  Although both edge
types are polar, the polarity of edges parallel to the Mn ($2 \times 1$) rows
could be smaller due to a better balance of ionic charges.  In the case
of the p($2 \times 2$) islands, the difference in polarity might be
smaller due to the additional top Mn ions in the vicinity of the
island edges.  Hence, island shapes close to quadratic are favored.
Islands with polar edge orientations that are immersed into the substrate
have also been observed for other rock system with rock salt structure, like CoO/Ag(001)
\cite{Shantyr2004b}, NiO/Ag(001) \cite{Sebastian1999}, MnO/Ag(001), and MgO/Ag(001) \cite{Ferrari2005}.  There,
the polar edge orientations have been found to be stabilized by island
immersion into the Ag substrate, which is enabled by the facile
diffusion of Ag even at room temperature. Similar immersion effects are
also expected for Mn$_3$O$_4$(001) islands of Fig.  \ref{fig:STM1} ,
in line with the presence of polar edges.

\subsection{Thin films of 6 and 12 sublayers}

\begin{figure}[ht]
 \includegraphics[width=\textwidth]{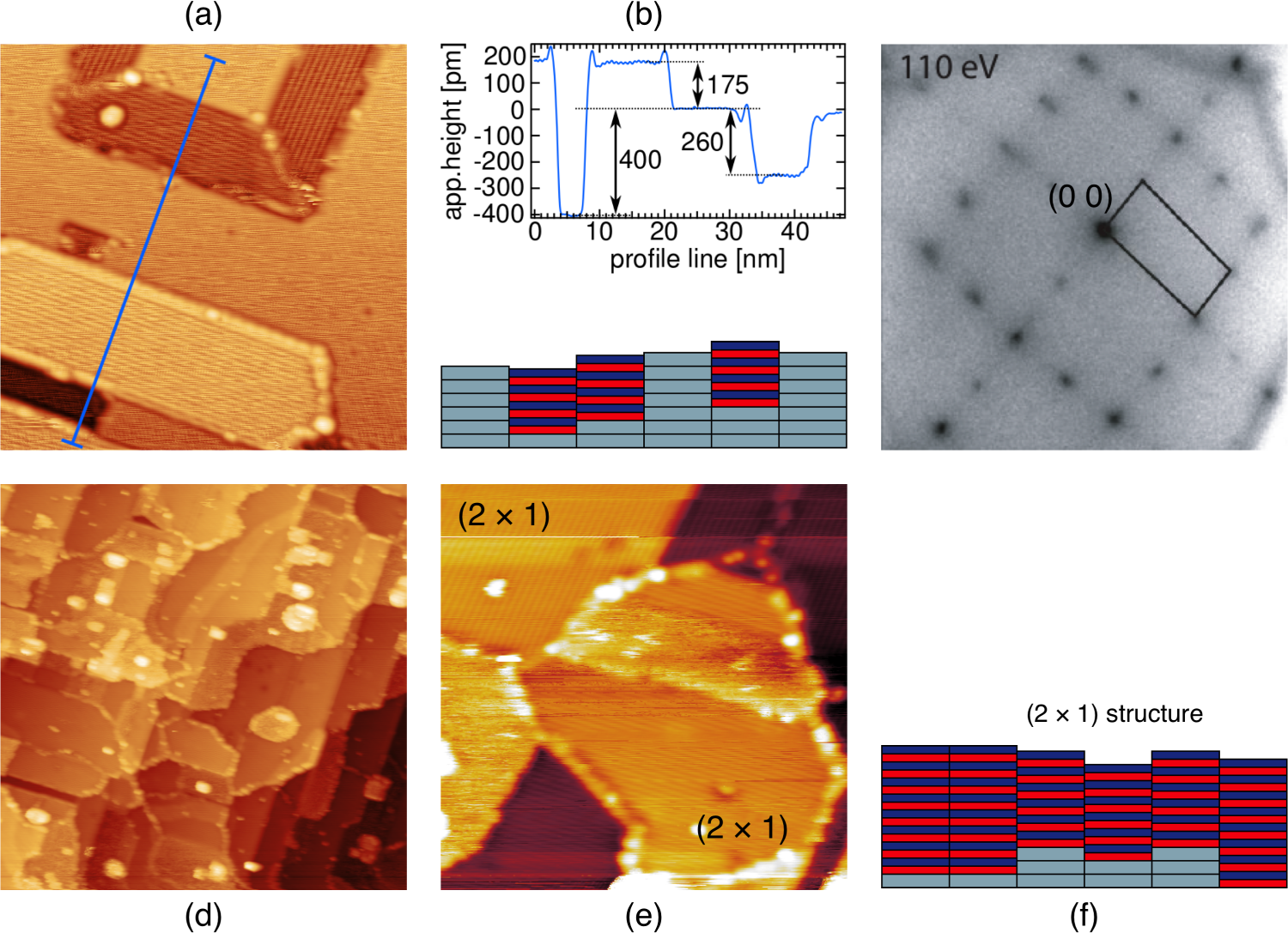}
  \caption{(a) STM image of a 6~sublayer Mn$_3$O$_4$
    film on Ag(001) (46 $\times$ 46~nm$^2$, 2.0~V, 0.8~nA) (b) height
		profile along the blue line; (c) LEED pattern of the oxide films
		of 6~sublayers with a p($2 \times 1$) unit cell.  (d) STM image of
		a 12~sublayer film Mn$_3$O$_4$ on Ag(001) (120 $\times$
		120~nm$^2$, 1.3~V, 0.8~nA) (e) zoomed-in STM image (36 $\times$
		36~nm$^2$, 2.0~V, 0.5~nA); (f) scheme of the 12~sublayer film on
		the silver substrate.
  }\label{fig:STM2} 
\end{figure}

Figures \ref{fig:STM2} a and d show STM images obtained after the
deposition of 6 and 12 sublayers of Mn$_3$O$_4$ on Ag(001).  After
deposition at room temperature the film was annealed to 630~K to
increase film quality. 
The STM image in Fig.  \ref{fig:STM2}a shows Mn$_3$O$_4$ islands with
p($2 \times 1$) stripes surrounded by flat Ag areas.  Whereas the
nominal thickness is 6~sublayers, the Mn$_3$O$_4$ islands cover only
70\% of the image.  Therefore, the local island thickness must be
considerably larger.  Different step heights can be found in the line
profile of Fig.  \ref{fig:STM2}b.  A height of 400~pm fits that of the
Ag(001) unit cell ($a_{Ag}$: 409~pm).  The other values of 260 and
175~pm, respectively, could result from Mn$_3$O$_4$ islands that are
embedded into the silver substrate as indicated by the scheme of Fig.
\ref{fig:STM2}b.  Due to different local thicknesses and/or depths of
embedding, islands appear with different contrasts.  However, no clear
conclusion about this fact is possible since the electronic structure
of the Mn$_3$O$_4$ islands needs to be taken into account when
determining heights from STM images \cite {Shantyr2004b}.  Exactly as
at low coverage, the p($2 \times 1$) islands prefer a rectangular shape
with long edges running along the MnO ($2 \times 1$) rows (compare to
Fig.  \ref{fig:STM1}).  The same islands are also observed but rotated
by 90$^\circ$, which is actually expected from the fourfold symmetry
of the Ag(001) substrate.  Where such islands meet, they form domain
boundaries along [100] directions.  Contrary to the initial stages
p($2 \times 2$) structures like the ones in Fig.  \ref{fig:STM1} a have
not been found for the 6~sublayer film.  The surface termination of
these p($2 \times 2$) structures is with Mn ions only and theoretical
investigations \cite{Bayer2007} have shown that for an Mn$_3$O$_4$
film with a thickness $>$8~sublayers the energy of this termination is
significantly higher than that of an Mn$_2$O$_4$ termination with its
p($2 \times 1$) structure.  The LEED pattern shown in Fig.
\ref{fig:STM2}c gives informations about the integral structure of the
prepared sample.  One can recognize a p($2 \times 1$) superstructure
with two domains that are rotated by 90$^\circ$.  In agreement to the STM
images, no spots of a p($2 \times 2$) superstructure have been found
at any electron energy.  The LEED pattern of the film of 12~sublayers
has been found to be identical to the 6~sublayer one.

\subsection{Thin films of more than 15 sublayers}

\begin{figure}
  \includegraphics[width=0.7\textwidth]{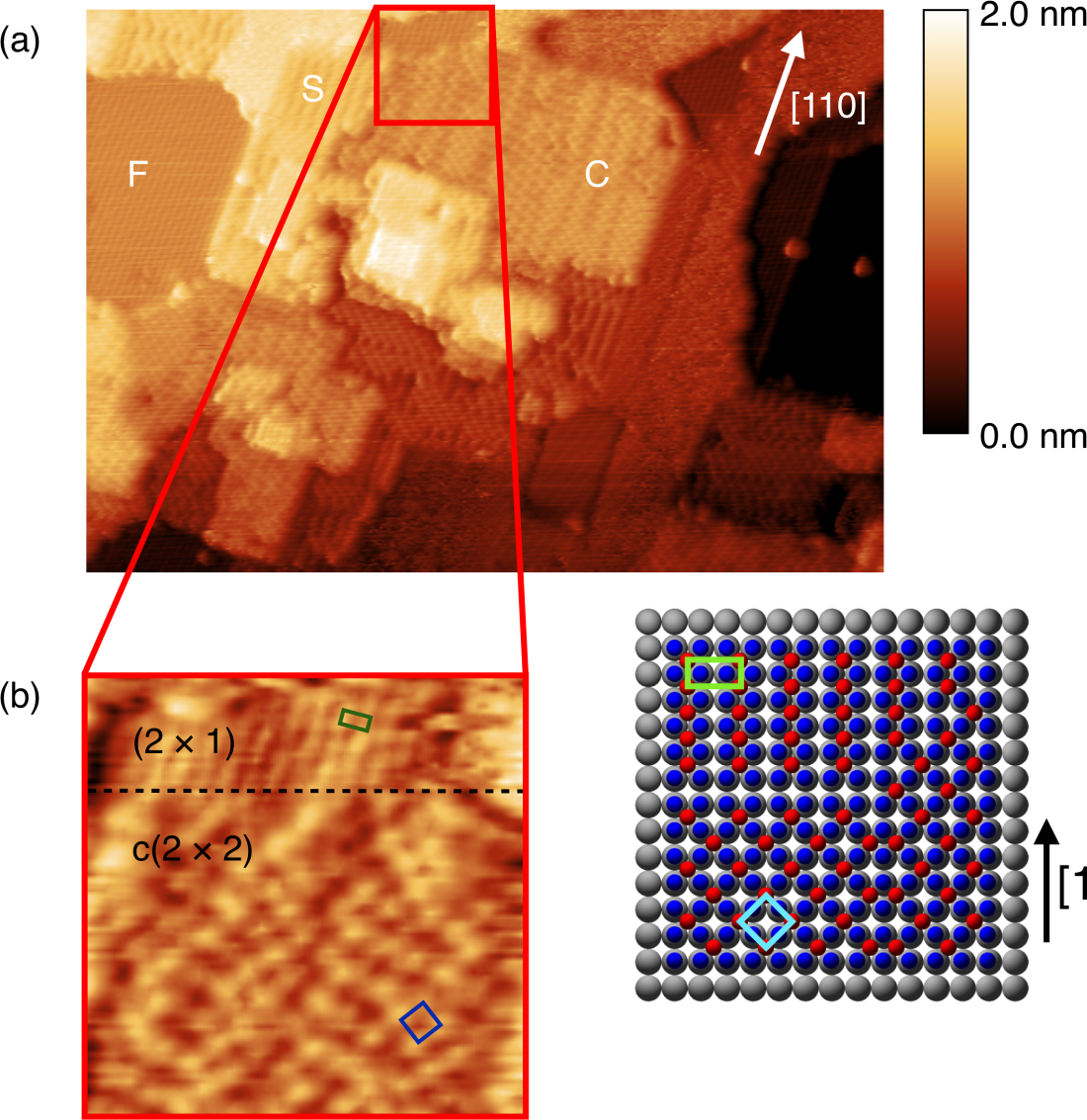}
  \caption{(a) STM image of a 15~sublayers
  Mn$_3$O$_4$ film on Ag(001) (58 $\times$ 47~nm$^2$, 2.0~V, 0.3~nA).
  The capital letters mark different structures at the surface; (b)
  STM detail (11$\times$11~nm$^2$, FFT filtered) of (a) together with
  the atomic model visualizing the phase transition from a p($2 \times
  1$) to a c($2 \times 2$) structure.  For further details see the text.
  }\label{fig:STM3} 
\end{figure}

\begin{figure}
  \includegraphics[width=0.7\textwidth]{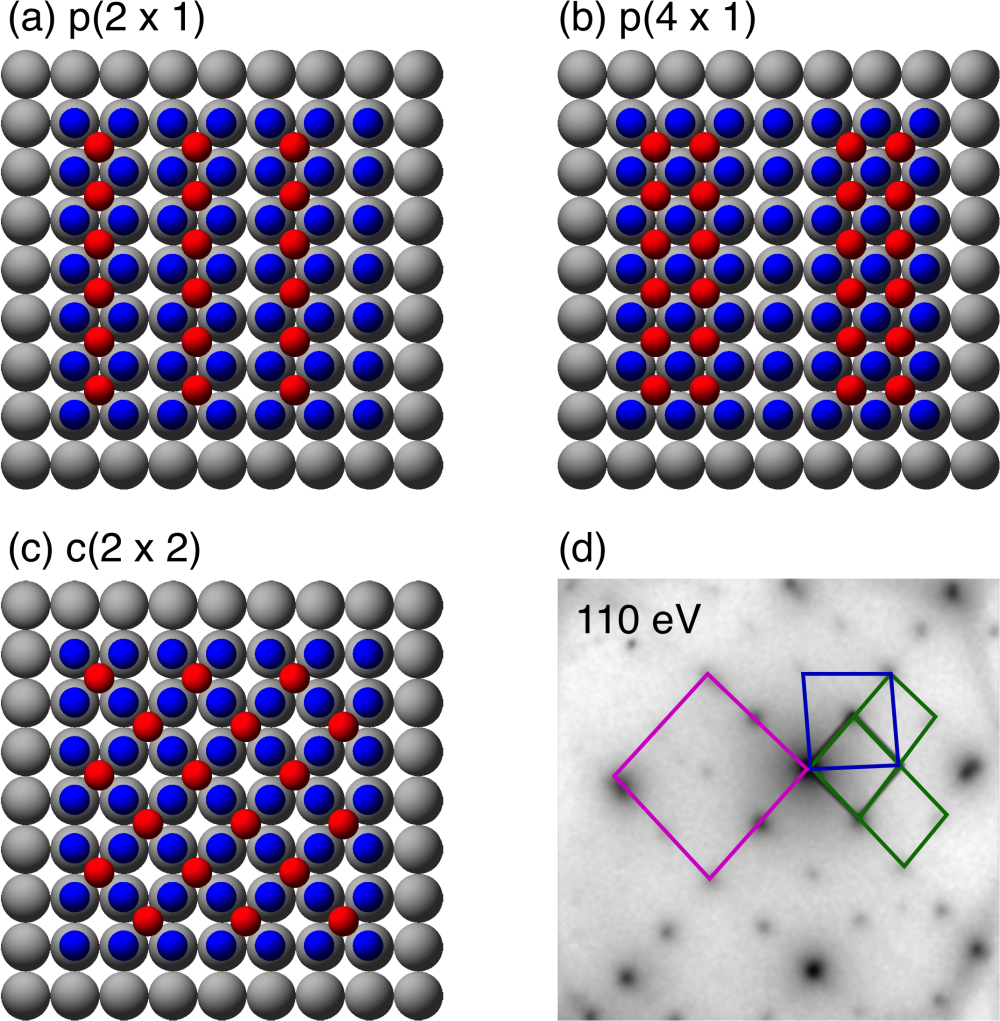}
  \caption{(a) p($2 \times 1$), (b) p($4 \times
    1$) and (c) c($2 \times 2$) Mn$_2$O$_4$ superstructures on Ag(001)
		and (d) LEED pattern at 110~eV for the 15 sublayer Mn$_3$O$_4$
		film on Ag(001) including unit cells of the p($2 \times 1$)
		(green), c($2 \times 2$) (blue), and p($2 \times 2$) (magenta)
		superstructures.
		}\label{fig:Structures} 
\end{figure}

After the deposition of 15~sublayers of Mn$_3$O$_4$ at room
temperature and subsequent annealing to 630~K one can distinguish five
different structures in the STM image of Fig.  \ref{fig:STM3}a.  The
flat regions F show the same stripe p($2 \times 1$) superstructure
like the islands found for 12 sublayers.  Additionally, stripe-like ($4
\times 1$) structures (S) and atomically resolved c($2 \times 2$)
structures (C) were found.  The wide stripes of the ($4 \times 1$)
superstructure are exactly twice as wide as those of the ($2 \times
1$) structure.  The $\times 1$ periodicity along the ($4 \times 1$)
stripes is not
resolved here.  Atomic models of the structures are
depicted in Fig.  \ref{fig:Structures}.  Starting point is the
Mn$_2$O$_4$ top layer from Fig.  \ref{fig:STM1}c.  The proposal for the ($4 \times 1$)
superstructure with wide stripes is obtained by moving every second Mn
row in the [110] direction towards a neighboring row (Fig.
\ref{fig:Structures}b).  The c($2 \times 2$) structure is obtained by
moving every second ion of an Mn row of the p($2 \times 1$) structure
along the [110] direction to the neighboring fourfold hollow site.
The resulting atomic pattern is rotated by 45$^\circ$ with respect to
the p($2 \times 1$) structure and the island edges become non-polar
(Fig.  \ref{fig:Structures}c).

The regions S with the striped ($4 \times 1$) superstructure in Fig. \ref{fig:STM3}a are much
smaller than the other ones.  Therefore, they do not result in
additional spots in the LEED pattern.  On the other hand, the number of
islands with a c($2 \times 2$) structure is large enough to lead to
additional spots in the diffraction pattern as is seen in Fig. \ref{fig:Structures}d.
The LEED pattern in Fig.  \ref{fig:Structures}d could result from two
different structural arrangements.  The first one is a p($2 \times 2$)
superstructure as found for the initial stages (Fig.
\ref{fig:STM1}a).  The second one is a combination of p($2 \times 1$)
and c($2 \times 2$) superstructures (see Fig.  \ref{fig:Structures}d).
Again, the p($2 \times 2$) superstructure is excluded because of the
substantially higher surface energy of the pure Mn termination
\cite{Bayer2007}.  Furthermore, the model with a mix of p($2 \times
1$) and c($2 \times 2$) superstructures is supported by the STM
images.  It is also supported by the fact that both the p($2 \times
1$) and c($2 \times 2$) superstructures correspond to a coverage of
0.5, which makes their coexistence and the transition from one to the
other very easy, whereas a p($2 \times 2$) superstructure usually
corresponds to a coverage of 0.25 or 0.75.  Therefore, transitions
between the p($2 \times 1$) and the p($2 \times 2$) superstructure
would require considerable mass transport across the surface.

The coexistence of the p($2 \times 1$) and c($2 \times 2$) structures can nicely
be seen in Fig.  \ref{fig:STM3}b.  One can recognize the p($2 \times
1$) rows in the upper part of the image and in the atomic model next
to the image.  At a certain point, marked with the dashed black line,
the rows begin to break down and the Mn ions are rearranged.  The
zigzag pattern of the c($2 \times 2$) mesh is visible in the model as
well as in the STM image.  Also, the rotation of the c($2 \times 2$)
pattern by 45$^\circ$ becomes obvious.

The coexistence of p($2 \times 1$) and c($2 \times 2$) structures
instead of a simple p($2 \times 2$) structure can also be underpinned
by the following consideration.  The p($2 \times 1$) and c($2 \times
2$) superstructures correspond to a coverage of 0.5, which makes their
coexistence and the transition from one to the other very easy,
whereas a p($2 \times 2$) superstructure usually corresponds to a
coverage of 0.25 or 0.75.  Therefore, transitions between the p($2
\times 1$) and the p($2 \times 2$) superstructure would require
considerable mass transport across the surface.

The driving force for the structural transition in the Mn$_2$O$_4$ top
layer is supposedly a decrease in surface energy.  Accompanied with
growing expansion and perfection of the c($2 \times 2$) superstructure
the amount of polar-edged p($2 \times 1$) islands decreases.
Consequently, the densities of rotated p($2 \times 1$) domains and
domain walls are expected to decrease.  This leads to a lower
interface energy for the c($2 \times 2$) superstructure.

\begin{figure}
  \includegraphics[width=\textwidth]{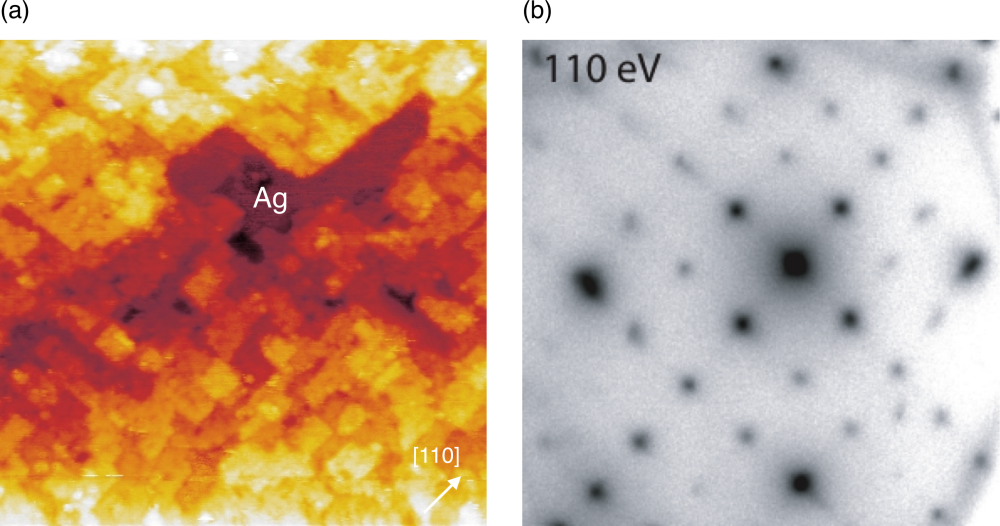}
  \caption{ STM image and LEED pattern of 28
  sublayers of Mn$_3$O$_4$ on Ag(110) after annealing to 740~K. The
  LEED pattern shows the superposition of a p($2 \times 1$) and c($2
  \times 2$) superstructure.  In the STM image, the edges of the
  Mn$_3$O$_4$ islands are oriented along [110] directions.  The large,
  deep holes with flat bottoms are assigned to the bare Ag(001) substrate.
  }\label{fig:LEED&STM}
\end{figure}

In order to test this assumption, 28~sublayers of Mn$_3$O$_4$ have
been deposited on Ag(001).  The increased thickness of the film allows
annealing up to 740~K. Although the film ruptures and a few areas with
the bare Ag(001) substrate become visible in STM (Fig.
\ref{fig:LEED&STM}b), the brilliance of the diffraction pattern and
the small size of the LEED spots indicate that film ordering and
perfection of the Mn$_3$O$_4$ islands are improved.  Along with this,
the orientation of island edges along [110] directions is more
pronounced in the STM image compared to lower coverages.  Despite the
high film thickness and the large annealing temperature, the LEED
pattern does not only show the expected c($2 \times 2$)
superstructure, but the p($2 \times 1$) superstructure is still
present (Fig.  \ref{fig:LEED&STM}b).  Obviously, the energy difference
between the p($2 \times 1$) and c($2 \times 2$) superstructures is small
even at high coverages of Mn$_3$O$_4$.  Therefore, the larger entropy
favors the simultaneous presence of two superstructures.  To
understand the mechanisms that lead to the coexistence of p($2 \times
1$) and c($2 \times 2$) structures even at thick Mn$_3$O$_4$ films, we
are looking forward to some theoretical descriptions.

\section{Conclusions}
The growth of ultrathin Mn$_3$O$_4$(001) films on Ag(001) has been
investigated from initial stages up to coverages of 28~sublayers (3.5 unit cells) by a
combination of near-edge X-ray absorption fine structure
spectroscopy, scanning tunneling microscopy, and low
energy electron diffraction.  For the initial stages the
Mn$_3$O$_4$ islands exhibit p($2 \times 1$) as well as p($2 \times 2$)
superstructures with island edges aligned along the [110] directions.  For 
increased film thickness up to 12~sublayers, Mn$_3$O$_4$ islands grow
embedded into the silver substrate having a p($2 \times 1$)
superstructures only.  Further increase of the film thickness leads to
a structural transition of the Mn$_3$O$_4$ film 
with a dominating c($2 \times 2$) superstructure next to the
p($2 \times 1$) superstructure.

\begin{acknowledgments}
Financial support from SFB~762, ``Functionality of Oxide Interfaces''
is gratefully acknowledged.  The authors
acknowledge the mutual technical support within the SFB and the
beamline staff at BESSY II, Berlin. Particular thanks goes to R.
Kulla for technical assistance as well as the scientific 
discussions in the SFB.
\end{acknowledgments}

%

\end{document}